\documentclass[lettersize,journal]{IEEEtran}
\usepackage{amsmath,amsfonts}
\usepackage{algorithmic}
\usepackage{algorithm}
\usepackage{array}
\usepackage{textcomp}
\usepackage{stfloats}
\usepackage{url}
\usepackage{verbatim}
\usepackage{graphicx}
\usepackage{cite}
\usepackage{siunitx}
\usepackage{multirow}
\usepackage{booktabs}
\usepackage{amssymb}
\usepackage{xcolor}
\usepackage{makecell}
\usepackage{threeparttable}
\usepackage[colorlinks,linkcolor=black,urlcolor=black,citecolor=black]{hyperref}
\begin{document}

\title{AudioSetCaps: An Enriched Audio-Caption Dataset using Automated Generation Pipeline with \\Large Audio and Language Models}


\author{Jisheng Bai,~\IEEEmembership{Graduated~Student~Member,~IEEE,}
	Haohe Liu,~\IEEEmembership{Graduated~Student~Member,~IEEE,}\\
	Mou Wang,~\IEEEmembership{Member,~IEEE,}
    Dongyuan Shi,~\IEEEmembership{Senior Member,~IEEE,}
    Wenwu Wang,~\IEEEmembership{Senior Member,~IEEE,}\\
    Mark D. Plumbley,~\IEEEmembership{Fellow,~IEEE,}
    Woon-Seng Gan,~\IEEEmembership{Senior Member,~IEEE,}
    Jianfeng Chen,~\IEEEmembership{Senior Member,~IEEE,}
	\thanks{This work was partly supported by the China Scholarship Council during a visit of Jisheng Bai to Nanyang Technological University.
    This research was partly supported by Engineering and Physical Sciences Research Council (EPSRC) under Grant EP/T019751/1 and EP/Y028805/1, a Research Gift from Adobe, and a PhD scholarship from the Centre for Vision, Speech and Signal Processing (CVSSP) at the University of Surrey and BBC R\&D. \textit{(Corresponding author: Jianfeng Chen.)} }
    \thanks{Jisheng Bai, Dongyuan Shi, and Jianfeng Chen are with the School of Marine Science and Technology, Northwestern Polytechnical University, Xi’an, China. Email: baijs@mail.nwpu.edu.cn; dongyuan.shi@nwpu.edu.cn; chenjf@nwpu.edu.cn)}
    \thanks{Haohe Liu, Wenwu Wang, and Mark D. Plumbley are with the Centre for Vision, Speech and Signal Processing (CVSSP), University of Surrey, Guilford, UK. Email: \{haohe.liu, w.wang, m.plumbley\}@surrey.ac.uk.}
    \thanks{Mou Wang is with the Institute of Acoustics, Chinese Academy of Sciences, Beijing, China. Email: wangmou21@mail.nwpu.edu.cn.}
    \thanks{Woon-seng Gan is with School of Electrical \& Electronic Engineering, Nanyang Technological University, Singapore. Email: ewsgan@ntu.edu.sg.}
    
}


\maketitle
\pagenumbering{arabic}


\begin{abstract} 
With the emergence of audio-language models, constructing large-scale paired audio-language datasets has become essential yet challenging for model development, primarily due to the time-intensive and labour-heavy demands involved.
While large language models (LLMs) have improved the efficiency of synthetic audio caption generation, current approaches struggle to effectively extract and incorporate detailed audio information.
In this paper, we propose an automated pipeline that integrates audio-language models for fine-grained content extraction, LLMs for synthetic caption generation, and a contrastive
language-audio pretraining (CLAP) model-based refinement process to improve the quality of captions. 
Specifically, we employ prompt chaining techniques in the content extraction stage to obtain accurate and fine-grained audio information, while we use the refinement process to mitigate potential hallucinations in the generated captions.
Leveraging the AudioSet dataset and the proposed approach, we create AudioSetCaps, a dataset comprising $1.9$ million audio-caption pairs, the largest audio-caption dataset at the time of writing. 
The models trained with AudioSetCaps achieve state-of-the-art performance on audio-text retrieval with R@1 scores of $46.3\%$ for text-to-audio and $59.7\%$ for audio-to-text retrieval and automated audio captioning with the CIDEr score of $84.8$. 
As our approach has shown promising results with AudioSetCaps, we create another dataset containing $4.1$ million synthetic audio-language pairs based on the Youtube-8M and VGGSound datasets.
To facilitate research in audio-language learning, we have made our pipeline, datasets with $6$ million audio-language pairs, and pre-trained models publicly available at \url{https://github.com/JishengBai/AudioSetCaps}.


\end{abstract}





\IEEEpeerreviewmaketitle




\section{Introduction}
\label{sec:into}



Audio-language learning plays a pivotal role in audio perception by enabling machines to process and understand relationships between audio signals and textual data~\cite{mei2024wavcaps}. 
This cross-modal understanding and processing ability is essential for machines to interpret, describe, and interact with the rich acoustic world in natural language. 
Recent advances in joint learning of audio and language representations have significantly enhanced audio understanding and reasoning capabilities~\cite{deshmukh2023pengi, zhao2023survey}.
These advances have led to substantial progress in various applications, including speech recognition~\cite{fathullah2024prompting}, music composition~\cite{liu2023audioldm, liu2024audioldm2}, and cross-modal retrieval~\cite{koepke2022audio, bai2024audiolog}, moving beyond traditional audio processing methods towards more comprehensive audio understanding.

Audio-language models (ALMs) combine audio and language pre-trained models to enable comprehensive representation learning through audio-text pairs~\cite{yang2024air, latif2023sparks, wijngaard2024audio}. 
A representative model is the contrastive language-audio pretraining (CLAP), which incorporates a contrastive learning framework to map audio and text into a shared multimodal space~\cite{wu2023large, elizalde2023clap}. 
This approach has demonstrated strong performance on various downstream tasks such as audio-text retrieval (ATR)~\cite{wu2023large}, zero-shot emotion recognition, and zero-shot music genre classification~\cite{elizalde2023clap}.
While these advances show the potential of ALMs in audio understanding, their further development is heavily dependent on large-scale paired audio-language datasets. 
Constructing such datasets presents significant challenges due to the considerable time and labour investments required, creating a critical bottleneck in advancing this field~\cite{mei2024wavcaps}.

To address the challenge of constructing large-scale audio-caption datasets, recent works have leveraged large language models (LLMs) for automated audio caption generation~\cite{zhao2023survey}. 
LAION-Audio-630K\cite{wu2023large} uses sentence templates to transform web-collected tags into captions, while WavCaps~\cite{mei2024wavcaps} employs ChatGPT to refine human-annotated audio descriptions into structured captions. 
Although these approaches facilitate audio-caption dataset construction, they have limitations in introducing more detailed acoustic content, which has been shown beneficial for improving the quality of synthetic audio captions~\cite{manivannan2024emotioncaps}.
Auto-ACD~\cite{sun2024auto} and Sound-VECaps~\cite{yuan2024improving} attempt to enrich acoustic caption by integrating audio-visual information. 
However, the requirement of paired audio-visual data increases the pipeline complexity and limits its scalability to audio-only data.

Recently proposed large audio-language models (LALMs) have shown promising abilities in extracting specific audio characteristics~\cite{deshmukh2023pengi, gong2024listen, gong2023joint, chu2023qwen}. 
LALMs advance beyond previous ALMs by integrating pre-trained audio models with LLMs for audio understanding. Pengi pioneers this direction by framing audio tasks as text generation ~\cite{deshmukh2023pengi}, LTU enhances the framework through diverse audio question-answer (QA) pairs~\cite{gong2024listen}. Qwen-Audio demonstrates strong capabilities in multi-turn dialogues and various downstream tasks by modelling audio and language information~\cite{chu2023qwen}. 
While these LALMs have achieved progress, there remains room for improvement in extracting and comprehending complex audio signals~\cite{yang2024air}. 
This suggests an opportunity to leverage the complementary strengths of LALMs in audio content extraction and LLMs in text generation for creating high-quality audio-language datasets.

We propose an automatic pipeline integrating LALMs, LLMs, and CLAP models to generate enriched audio-language data. 
This pipeline consists of three main stages: LALMs-driven audio content extraction, LLMs-assisted caption generation, and CLAP model-based caption refinement.
In the first stage, we employ LALMs with a prompt chaining method to extract audio content, including overall and fine-grained audio information such as spoken language, speech emotion, music genre, and musical instruments. 
We then utilize LLMs to transform these audio contents into natural, detailed captions. 
In the final stage, we design a caption refinement module incorporating the CLAP model to filter out low-quality captions and mitigate LLM hallucinations.
Using this pipeline, we generated AudioSetCaps, a dataset containing over $1.9$ million~(M) captions for recordings in AudioSet~\cite{gemmeke2017audio}. 
Subjective evaluation shows that our dataset achieves higher caption quality than existing LLMs-based datasets. 
We conduct experiments on multiple audio-language tasks, such as ATR, zero-shot classification, and automated audio captioning (AAC).
The models trained on our dataset benefit from dataset scaling and achieve state-of-the-art performance.

Our main contributions are summarized as follows:
\begin{itemize}
\item[$\bullet$] We propose an automated pipeline that integrates LALMs and LLMs for audio-language dataset construction. Through prompt chaining techniques and refined caption generation, our pipeline can extract fine-grained audio content and iteratively improve the quality of captions.
\item[$\bullet$] We create AudioSetCaps, a dataset comprising $1.9$ M audio-caption pairs. Subjective mean opinion scores indicate that the captions in AudioSetCaps are of competitive quality. Experimental results demonstrate that models trained on AudioSetCaps achieve state-of-the-art performance in audio-language tasks, surpassing the models trained on previous datasets generated using LLMs-based pipelines.
\item[$\bullet$] 
We extend our pipeline to generate over $6$ M audio-caption pairs across multiple datasets, creating the largest audio-caption dataset at the time of writing.
Our pipeline, built with open-source models and tools, can run on consumer-grade GPUs, making it widely accessible. All resources related to this paper, including the proposed automatic audio labelling pipeline, datasets, and pre-trained models, are publicly available to facilitate future research in audio-language learning.
\end{itemize}

The remainder of this paper is organized as follows. 
Section \ref{sec:Related Work} introduces related works in audio-language learning. 
Section \ref{sec:method} introduces the proposed pipeline of dataset collection and statistics of the AudioSetCaps dataset. 
Section \ref{sec:exps} introduces experiments on audio-language learning tasks. 
Section \ref{sec:Conclusion} concludes this paper.

\section{Related Work}
\label{sec:Related Work}

Audio-language datasets are important in advancing audio-language learning, serving as the cornerstone for developing robust ALMs. 
In audio captioning, current datasets can be categorized into two main types: human-annotated datasets and those generated through language model-assisted pipelines.

\noindent
\textbf{Audio caption datasets with human annotation} form the foundation for developing and evaluating early audio captioning models. Through careful human annotation processes, experts create high-quality pairings of audio samples with descriptive text~\cite{kim2019audiocaps, drossos2020clotho, morato2021diversity}. This approach ensures precise and contextually relevant descriptions, though it faces scalability challenges due to the labour-intensive nature of manual annotation.
Two prominent datasets in this field are AudioCaps and Clotho. 
AudioCaps~\cite{kim2019audiocaps} is built upon AudioSet, consisting of $46,000$ YouTube audio clips, focusing on $75$ sound categories, where annotations were collected through Amazon Mechanical Turk\footnote{[Online]. Available: \url{https://requester.mturk.com/}}. AudioSet labels and corresponding video are provided as optional references during annotation to facilitate consistent and high-quality annotations. 
Clotho~\cite{drossos2020clotho} contains $4,981$ carefully curated audio samples from the Freesound platform, where each sample is annotated with five captions collected through workers on Amazon Mechanical Turk. The annotation process of the Clotho dataset only has access to the audio content without any additional context or hints. 
This makes sure the annotation is purely audio-dependent and captures different human perceptions of sounds.

\noindent
\textbf{Audio caption datasets with language model-assisted generation pipelines} aim to address the time-consuming and labour-intensive challenge associated with human annotation. 
By leveraging advanced language models, these generation pipelines can efficiently construct large-scale audio-text paired datasets. 
In this case, the annotation pipelines typically involve pre-trained language models to generate captions based on existing audio metadata, followed by filtering and other quality control mechanisms. 

Several notable datasets have been developed using language model-assisted generation pipelines. 
LAION-Audio-630K~\cite{wu2023large}, containing $633,526$ pairs, uses a pre-trained T5 model to generate captions for samples with only tags or label annotations. 
WavCaps~\cite{mei2024wavcaps}, comprising approximately $400,000$ audio-caption pairs, employs a three-stage processing pipeline with ChatGPT to filter and transform web-crawled descriptions into high-quality captions. 
Auto-ACD~\cite{sun2024auto}, with over $1.5$ million audio-text pairs, utilizes pre-trained models and APIs to extract audio and visual clues, then uses an LLM to organize various metadata into captions. 
Sound-VECaps~\cite{yuan2024improving} leverages an LLM to transform visual captions, audio captions, and tagging labels into descriptions, resulting in $1.66$ million synthetic audio-caption pairs. 
LAION-Audio-630K and WavCaps focus on transforming existing annotations but have limited fine-grained acoustic details.
Auto-ACD and Sound-VECaps not only increase the pipeline complexity but also require paired audio-visual data, limiting the scalability of caption generation.

\begin{figure*}[th]
\centering
\includegraphics[width=0.95\linewidth]{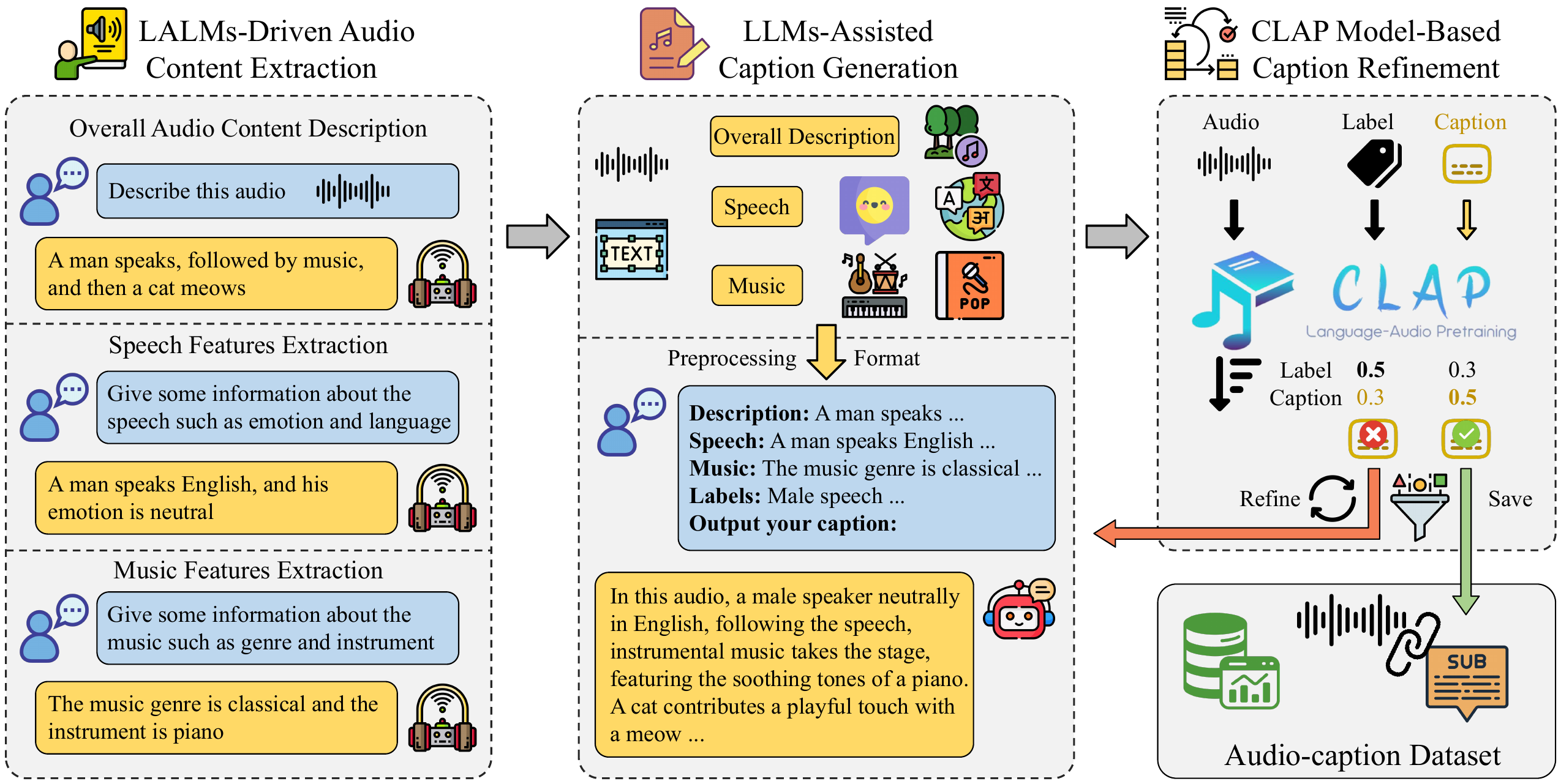}
\caption{Overview of the proposed automated caption generation pipeline, which integrates large audio-language models (LALMs), large language models (LLMs), and contrastive language-audio pretraining (CLAP) model.}
\label{fig:main_pipeline}
\end{figure*}

\section{AudioSetCaps Pipeline and Dataset}
\label{sec:method}
In this section, we present a comprehensive framework for automated audio caption generation and the resulting AudioSetCaps dataset. 
We first introduce the large models employed in our pipeline, explaining their capabilities and roles.
We then detail our three-stage pipeline, describing how each stage works together to generate enriched audio captions. Following this, we analyze the AudioSetCaps dataset in terms of its statistical characteristics and content distribution.
Finally, we validate the quality of our generated captions through human evaluation.

\subsection{Large Audio and Language Models in Pipeline}
\label{sec:large_models}
Our pipeline integrates three types of models, each selected for its specific strengths. 
The LALMs are used for content extraction, the LLMs handle text generation, and the CLAP model refines caption quality. 
This section introduces these models and describes their roles in our pipeline.

\noindent
\textbf{Qwen-Audio}~\cite{chu2023qwen} is an audio understanding LALM that supports various audio tasks.
Building upon Qwen-Audio, Qwen-Audio-Chat~\cite{chu2023qwen} can follow instructions, enable multi-turn dialogues, and support different downstream audio tasks.
Qwen-Audio sets a new benchmark across multiple audio tasks, achieving state-of-the-art performance on speech recognition, vocal sound classification, and acoustic scene classification~\cite{chu2023qwen}.
We use text prompts and audio recordings as inputs for Qwen-Audio-Chat, which generates descriptions of the overall audio content and detailed information about speech, and music.

\noindent
\textbf{Mistral 7B}~\cite{jiang2023mistral} is a $7.3$ billion parameter language model that demonstrates good performance across various benchmarks, outperforming larger models like Llama 2 13B~\cite{touvron2023llama2} and Llama 1 34B~\cite{touvron2023llama} in many areas.
Mistral 7B employs grouped-query and sliding window attention methods to improve the inference efficiency of long sequences, facilitating its deployment across various tasks. 
In our workflow, Mistral 7B transforms the extracted audio content into complete audio captions.

\noindent
\textbf{CLAP}~\cite{wu2023large, elizalde2023clap} models incorporate a contrastive learning framework to bring audio and text descriptions into a joint multimodal space, learning the mapping relationship between the two modalities.
Trained on LAION-Audio-630K, a large-scale audio caption dataset of 633,526 audio-text pairs, the LAION CLAP model~\cite{wu2023large} achieves state-of-the-art performance in text-to-audio retrieval and zero-shot audio classification.
CLAP models have been used to measure the semantic similarity between audio and text in many audio tasks~\cite{xiao2024reference, liu2024audioldm2}. 
Accordingly, we employ the LAION CLAP model~\cite{wu2023large} to evaluate the similarity between the generated captions and the corresponding audio recordings.

\subsection{Automated Audio-Caption Generation Pipeline}
\label{sec:pipeline}
The proposed automated pipeline comprises three key stages, the overview of the proposed pipeline is shown in Fig. \ref{fig:main_pipeline}.
Our pipeline begins with LALMs-driven audio content extraction, where we utilize LALMs to comprehensively analyze and extract different audio metadata from the input recordings. 
The extracted content feeds into the LLMs-assisted caption generation stage, where LLMs transform the detailed audio information into audio captions. 
Finally, the CLAP model-based caption refinement stage filters and refines the captions generated by LLMs, further improving the quality of the captions. 
This sequential approach allows us to take advantage of each model for efficient audio caption generation.

\begin{figure}[t]
\centering
\includegraphics[width=0.98\linewidth]{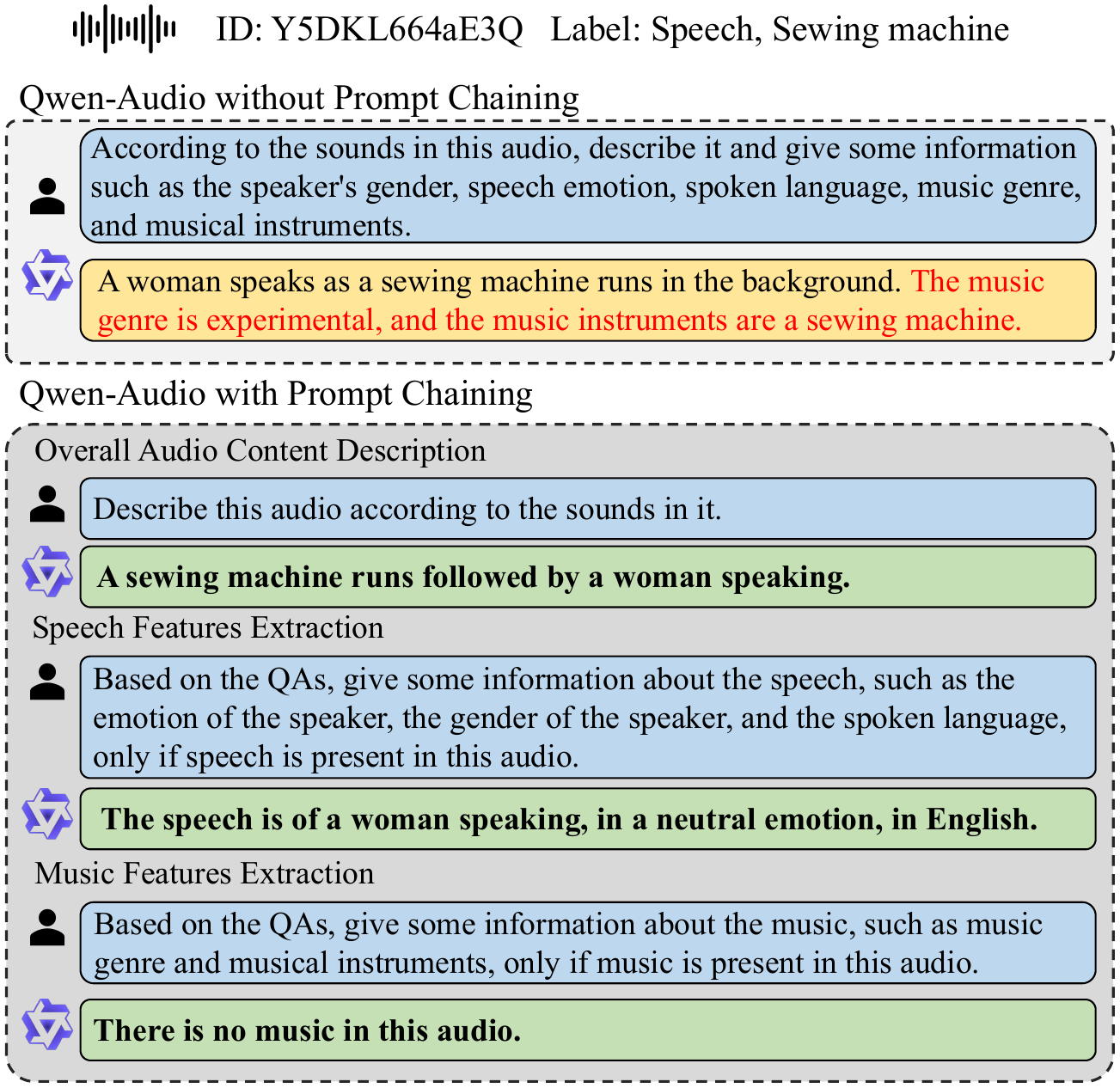}
\caption{An example of the output descriptions from Qwen-Audio with or without prompt chaining. Qwen-Audio without prompt chaining fails to output correct content in the second sentence (marked in red), Qwen-Audio with prompt chaining can output correct answers with fine-grained information.}
\label{fig:qwen_promptchaining}
\end{figure}

\subsubsection{LALMs-Driven Audio Content Extraction}
The primary goal of LALMs-driven audio content extraction is to obtain various fine-grained audio content from audio recordings. 
This approach extracts detailed information about speech, music, and environmental sounds to facilitate the audio-language learning of different acoustic representations. 
Using this enriched content is designed to enhance the learning of audio-language relationships and improve the performance of ALMs on downstream audio tasks, such as audio captioning~\cite{manivannan2024emotioncaps}.
However, direct audio content extraction using LALMs often results in incomplete or inconsistent information, especially for complex audio scenes containing multiple elements.

To address this challenge, we employ Qwen-Audio-Chat~\cite{chu2023qwen}, a multi-task pre-trained LALM, in combination with a prompt chaining technique. 
Prompt chaining~\cite{wu2022ai} is designed to break down complex tasks into smaller, more manageable sub-tasks, enabling the model to handle complex tasks, without requiring additional training or fine-tuning.
It has been applied in areas such as information extraction~\cite{kwak2024classify} and text summarization~\cite{sun2024prompt}.
Generally, our audio content extraction process consists of three main steps:
\begin{itemize}
\item \textit{Overall Audio Content Description}: Prompting Qwen-Audio to provide a concise overview of the audio, focusing on essential elements and their relationships.
\item \textit{Speech Features Extraction}: For detected speech, analyzing specific features such as emotion, gender, and language of the speaker.
\item \textit{Music Content Extraction}: For identified music, extracting information about genres and instruments.
\end{itemize}

This prompt chaining approach, adapted from LLM prompt engineering, is designed to enhance the extraction reliability of audio content by breaking down tasks into related subtasks. 
As shown in Fig.~\ref{fig:qwen_promptchaining}, without prompt chaining, Qwen-Audio may fail to output correct content or miss fine-grained information.

\begin{table}[t]
\centering
\caption{The prompts for LLMs-assisted caption generation.}
\label{tab:llms_prompt}
\begin{tabular}{p{0.95\columnwidth}}
\toprule
\textbf{Role Setting:} 
I'm currently doing crowd-sourced labeling of audio data. I want you to act as a professional worker to write the final caption for the audio. 
The most important thing is following my instructions! Now, let's get started. \\
\midrule
\textbf{Input Details:} \\
1. Crowd-sourced workers: \\
  \quad Description: \{\textit{Overall Audio Content Description}\} \\
  \quad Speech: \{\textit{Speech Features Description}\} \\
  \quad Music: \{\textit{Music Features Description}\} \\
2. The ground truth labels: \{\textit{ground truth labels}\} \\
\textbf{Instructions:} \\
- Do not mention the specific content of speech! \\
- Do not output the ground truth labels in the caption! \\
- Output your caption (within 50 words): \\ \bottomrule
\end{tabular}
\end{table}

\subsubsection{LLMs-Assisted Caption Generation}
The key challenge in caption generation is combining various types of extracted audio information while maintaining natural descriptions of audio content. 
In this stage, we leverage Mistral 7B~\cite{jiang2023mistral}, an LLM that can organize different audio information into audio captions using carefully designed prompts.

As shown in Fig. \ref{fig:main_pipeline}, we first implement a preprocessing step that refines the LALM-extracted content using regular expression matching. 
This process removes phrases that indicate the absence of certain audio elements (e.g., ``there is no speech/music present"), helping the LLM focus only on the audio content that is present. 
The refined content is then formatted into a structured prompt that integrates the information: overall audio content description, detailed speech features, and music features. 
To improve the reliability and accuracy of the generated captions, we include ground truth labels as references in the prompt, providing verified information about the audio content.

We design prompts with specific instructions to guide the LLM in generating appropriate captions, as shown in Table \ref{tab:llms_prompt}.
We prompt the LLM to play the role of a professional audio caption annotator and direct the LLM to avoid directly describing ground truth labels and mentioning specific speech recognition content in the output captions.
The prompts are designed to improve the quality of audio captions by leveraging the language processing strengths of LLMs.

\subsubsection{CLAP Model-Based Caption Refinement}

\begin{table}[t]
\centering
\caption{An example of audio caption before and after applying the CLAP model-based caption refinement. The refined caption removes the statements about absent elements of spoken language and music.}
\begin{tabular}{p{0.16\columnwidth}p{0.74\columnwidth}} 
\toprule
\textbf{Audio ID} & Y--0PQM4-hqg\\ \midrule
\textbf{Labels} & Gurgling, Waterfall, Stream \\ \midrule
\textbf{Before\quad Refinement} & A fast-moving body of water is featured prominently in this recording, with its distinct gurgling and rushing sounds creating a lively ambiance. No spoken language or musical elements are present within the audio.  \\ \midrule
\textbf{After\quad Refinement} & Fast-paced and powerful, this audio captures the dynamic flow of water in a river. The gentle yet forceful gurglings echo through the recording, evoking images of a rushing stream or perhaps the roar of a nearby waterfall.  \\ \bottomrule
\end{tabular}
\label{tab:audio_refinement}
\end{table}

Despite the strengths of LALMs and LLMs in extracting audio content and generating captions, the potential for hallucinations or inaccuracies in the generated captions remains a concern. 
Even with ground truth labels as a reference, LLMs may occasionally produce content that deviates from the actual audio information or includes extraneous details. 
To address this challenge, we leverage the LAION CLAP model~\cite{wu2023large}, which has demonstrated strong performance in assessing audio-text alignment through its training on extensive audio-language datasets.

In the refinement stage, we implement a quality control mechanism. 
We first use ground truth labels as quality benchmarks by computing the similarity scores between the audio and both the generated caption and the ground truth label. 
When the similarity score of a caption falls below the ground truth label score, indicating potential quality issues, we trigger the regeneration process. 
In addition, the number of regeneration attempts can be set to a certain number to maintain computational efficiency while applying the refinement.

Table \ref{tab:audio_refinement} shows the effectiveness of our refinement process through a concrete example. 
The refined caption eliminates redundant statements about absent elements while enhancing the description of actual acoustic content with more focused and vivid expressions. 
This improvement aligns with our goal of generating captions that accurately and efficiently describe the audio content without introducing irrelevant information.

\subsection{Dataset Statistics and Analysis}
\label{sec:dataset_statistics}
Using the proposed pipeline, we have generated AudioSetCaps, a large-scale audio-caption dataset derived from AudioSet recordings. 
In this subsection, we present a comprehensive analysis of AudioSetCaps and compare it with several popular audio caption datasets. 
Our analysis focuses on two complementary aspects: the overall statistics that demonstrate the scale and diversity of our dataset, and the fine-grained audio content that highlights its detailed acoustic characteristics.

\subsubsection{Overall Statistics}

\begin{table}[t]
\centering
\caption{The statistics comparison of several audio-caption datasets. Quan denotes the number of captions, Len represents the average caption length in words, and Voc indicates the vocabulary size. Caption generation methods include human annotation (H), audio model (A), visual model (V), and language model (L).}
\resizebox{0.49\textwidth}{!}{
\begin{tabular}{ccccc}
\toprule
\textbf{Dataset}                                 & \textbf{Quan}    & \textbf{Len}               & \textbf{Voc}   & \textbf{Source} \\ \midrule
Clotho \cite{drossos2020clotho}         & $30$~K      & $11$                 & $4$~K       & H     \\
AudioCaps \cite{kim2019audiocaps}       & $57$~K      & $9$                  & $5$~K       & H      \\ \midrule
LAION-Audio-630K \cite{wu2023large}     & $630$~K     & $7$                  & $311$~K     & L\\
WavCaps \cite{mei2024wavcaps}           & $400$~K     & $8$                  & $24$~K      & L      \\
Auto-ACD \cite{sun2024auto}             & $1.5$~M     & $18$                 & $20$~K      & A+V+L      \\
Sound-VECaps \cite{yuan2024improving}   & $1.6$~M     & $40$                 & $50$~K       & A+V+L      \\
AudioSetCaps                            & $1.9$~M     & $28$                 & $21$~K      & A+L      \\
\bottomrule
\end{tabular}
}
\label{tab:Dataset}
\end{table}

Table \ref{tab:Dataset} presents a comparison between AudioSetCaps and several popular audio caption datasets. AudioSetCaps contains $1.9$M audio-caption pairs derived from AudioSet recordings, surpassing the scale of compared datasets. The average caption length of AudioSetCaps is $28$ words, providing longer descriptions compared to LAION-Audio-630K ($7$ words) and WavCaps ($8$ words). 
Although AudioSetCaps has a relatively shorter average length ($28$ words) and smaller vocabulary size ($21$K) compared to Sound-VECaps, it focuses exclusively on audio content analysis by combining LALMs and LLMs.

\begin{figure*}[ht]
    \centering
    \begin{minipage}[b]{0.48\textwidth}
        \centering
        \includegraphics[width=\textwidth]{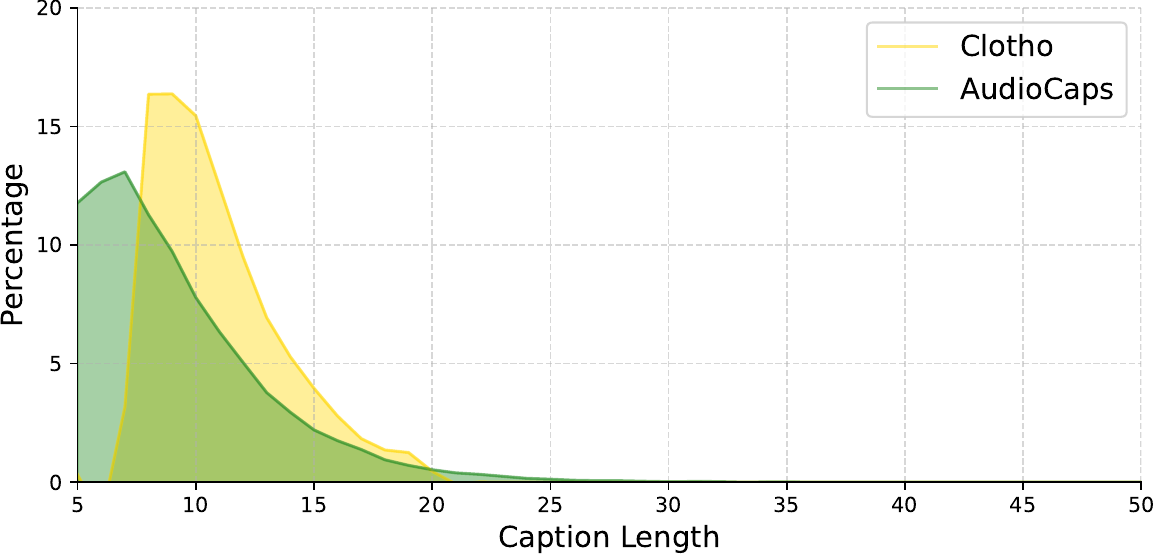}
    \end{minipage}
    \hfill
    \begin{minipage}[b]{0.48\textwidth}
        \centering
        \includegraphics[width=\textwidth]{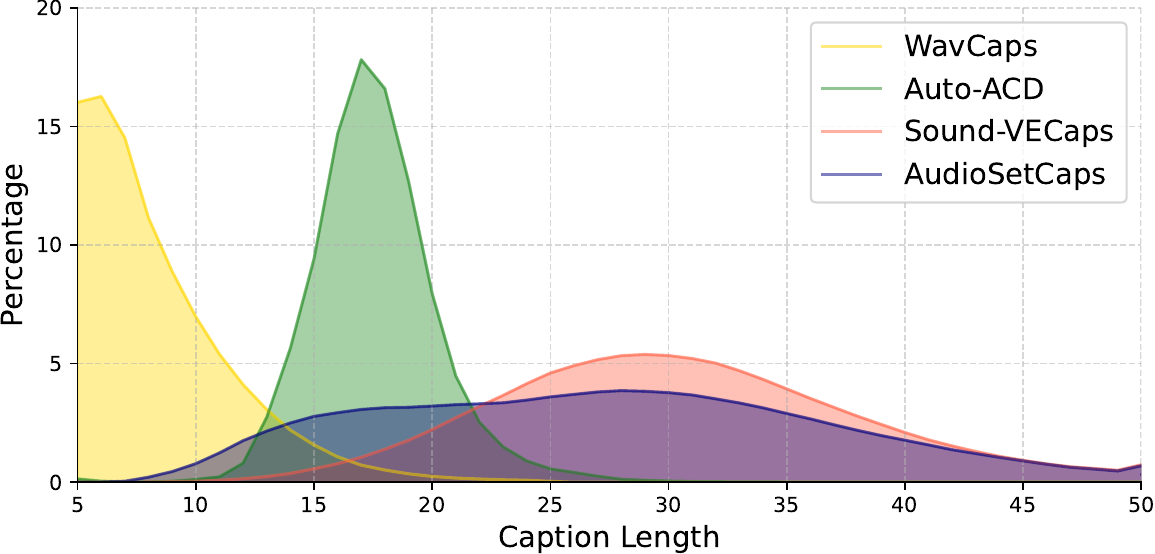}
    \end{minipage}
    \caption{Distribution of caption lengths across several popular audio caption datasets. The left subfigure shows the caption length distributions of human-labeled audio caption datasets, and the right subfigure shows the caption length distributions of LLMs-assisted audio caption datasets.}
    \label{fig:len_dis}
\end{figure*}

Fig. \ref{fig:len_dis} presents the distribution of caption lengths across different datasets, revealing distinct patterns between human annotations and automated approaches. 
Human-annotated datasets, i.e., Clotho and AudioCaps, exhibit concentrated distributions with most captions between $5$ to $15$ words, reflecting the tendency of human annotators towards concise descriptions. 
As for the distribution of automated approaches, WavCaps produces relatively short captions, and AudioSetCaps shows a more balanced distribution of caption length.
Auto-ACD, Sound-VECaps, and AudioSetCaps generate longer descriptions than WavCaps, indicating the incorporation of other modalities in the pipeline produces longer audio captions.

\subsubsection{Fine-grained Audio Information}

\begin{table}[t]
\centering
\caption{Classes used to collect fine-grained audio information.}
\begin{tabular}{p{0.25\columnwidth}p{0.6\columnwidth}}
\toprule
\textbf{Fine-grained Audio Information} & \textbf{Classes} \\
\midrule
Spoken Language & English, French, Chinese, Japanese, Arabic, Spanish, Hindi, Russian, German, Portuguese \\
Speech Emotion & Neutral, Calm, Angry, Excited, Sad, Happy, Fearful, Surprised, Frustrated, Nervous  \\
Music Instrument & Guitar, Bass, Piano, Drums, Violin, Flute, Trumpet, Saxophone, Clarinet, Harp \\
Music Genre & Electronic, Pop, Folk, Rock, Classical, Country, Jazz, Blues, Hip hop, Reggae \\ \bottomrule
\end{tabular}
\label{tab:audio_categories}
\end{table}


\begin{table}[t]
\centering
\caption{Statistics of different audio features for LLMs-assisted audio-caption datasets.}
\setlength{\tabcolsep}{3pt}
\begin{tabular}{lcccc}
\toprule
 & Language (K) & Emotion (K) & Instrument (K) & Genre (K) \\ \midrule
WavCaps & $0.01$ & $1.4$ & $11.4$ & $3.8$ \\
Auto-ACD & $3.5$ & $16.6$ & $299.8$ & $214.3$ \\
Sound-VECaps & $6.7$ & $64.3$ & $207.7$ & $134.1$ \\
AudioSetCaps & $13.8$ & $336.0$ & $690.0$ & $682.0$ \\
\bottomrule
\end{tabular}
\label{tab:speech_music_sta}
\end{table}

\begin{figure*}[t]
    \centering
    \begin{minipage}[b]{0.27\textwidth}
        \centering
        \includegraphics[width=\textwidth]{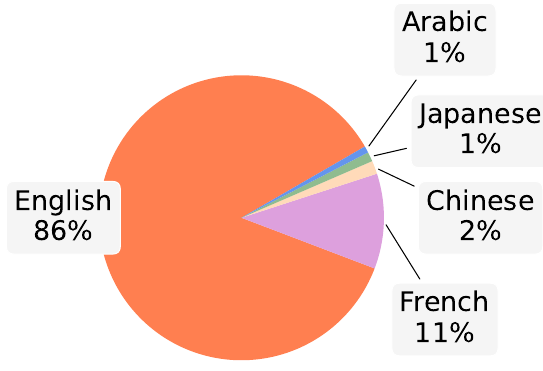}
    \end{minipage}
    \hfill
    \begin{minipage}[b]{0.24\textwidth}
        \centering
        \includegraphics[width=\textwidth]{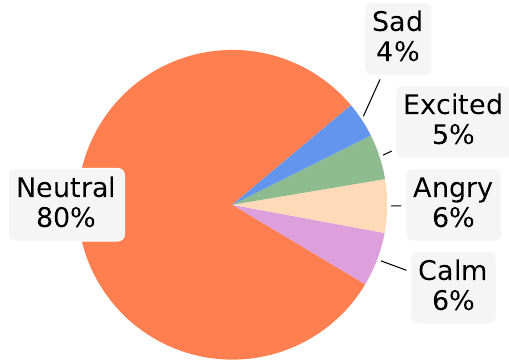}
    \end{minipage}
    \hfill
    \begin{minipage}[b]{0.19\textwidth}
        \centering
        \includegraphics[width=\textwidth]{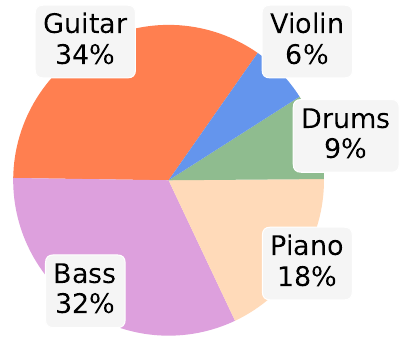}
    \end{minipage}
    \hfill
    \begin{minipage}[b]{0.21\textwidth}
        \centering
        \includegraphics[width=\textwidth]{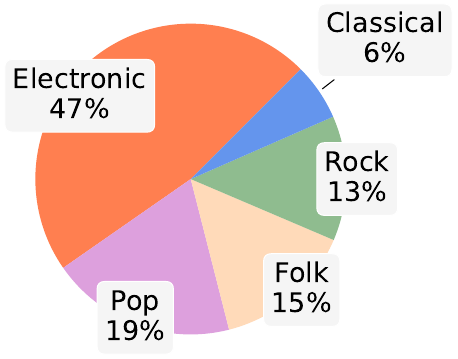}
    \end{minipage}
    \caption{Distribution of the top five categories of fine-grained speech and music information in AudioSetCaps, with pie charts representing the proportion of the most frequent categories in each type. The subfigures from left to right are pie charts of spoken languages, speech emotions, musical instruments, and music genres.}
    \label{fig:fine_grained_pie}
\end{figure*}

We count the statistics of fine-grained audio information in LLMs-based audio-language datasets to analyze the characteristics of audio-caption datasets. 
Table \ref{tab:audio_categories} shows the classes for collecting fine-grained audio information, including spoken languages, speech emotions, musical instruments, and music genres.
The statistical results are illustrated in Table \ref{tab:speech_music_sta}. 
AudioSetCaps reveals its comprehensive coverage of fine-grained audio content compared to other LLMs-based datasets, with consistently larger numbers of identified instances across all categories. 
In music genre recognition, AudioSetCaps identifies $682.0$~K instances, more than three times the number in Auto-ACD or Sound-VECaps.

The distribution analysis of fine-grained information in AudioSetCaps reveals distinct patterns of different categories, as illustrated in Fig. \ref{fig:fine_grained_pie}. In terms of languages, \textit{English} dominates with $86$\% of the instances, reflecting a significant bias in the dataset. Speech emotions are dominated by \textit{Neutral} with $80$\% of the instances. For musical instruments, \textit{Guitar} and \textit{Bass} are the most prevalent at $34$\% and $32$\%, respectively. Music genres show a notable preference for \textit{Electronic} music at $47$\%.

\begin{figure}[t]
    \centering
    \includegraphics[width=0.95\linewidth]{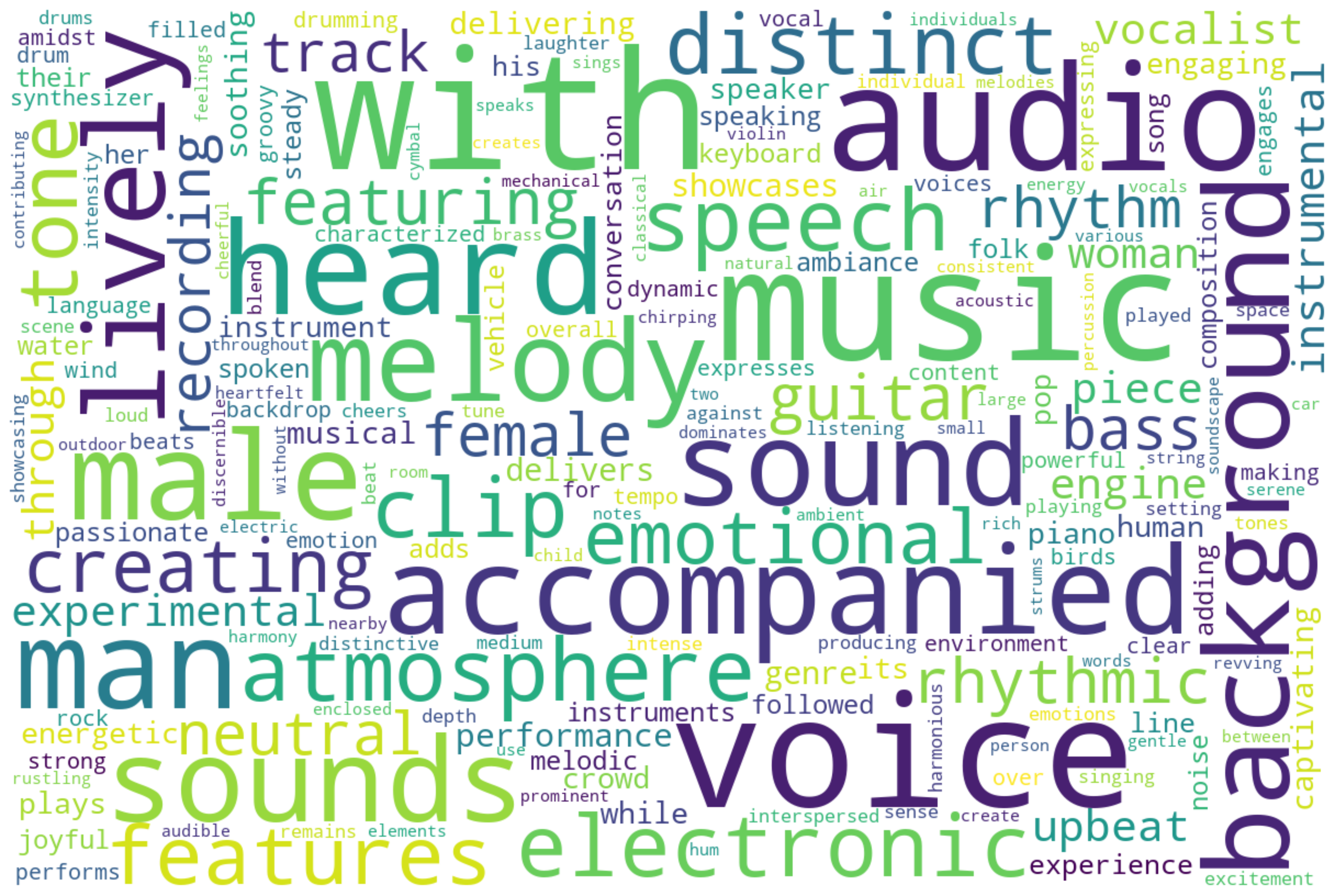}
    \caption{Caption word cloud of AudioSetCaps excluding stop words.}
    \label{fig:wordcloud}
\end{figure}

We provide a vocabulary word cloud of AudioSetCaps to show the rich and diverse content of the dataset, shown in Fig. \ref{fig:wordcloud}. 
Prominently featured are key audio elements such as ``voice", ``music", and ``sound", indicating the comprehensive coverage of various audio types. 
The presence of descriptions like ``female voice", ``speech", and ``neutral" suggests detailed annotations of speech characteristics. 
Musical elements are well-represented through instruments including ``guitar", ``bass", and ``piano" and genres such as ``electronic". 
The word cloud also reveals a focus on audio qualities and acoustic atmosphere, with descriptive terms like ``distinct", ``rhythmic", ``emotional", and ``accompanied" appearing frequently. 
This rich vocabulary demonstrates how AudioSetCaps captures both the acoustic content and its contextual characteristics.

\subsection{Subjective Evaluation}
\label{sec:sub_eval}

\begin{figure}[t]
    \centering
    \includegraphics[width=1\linewidth]{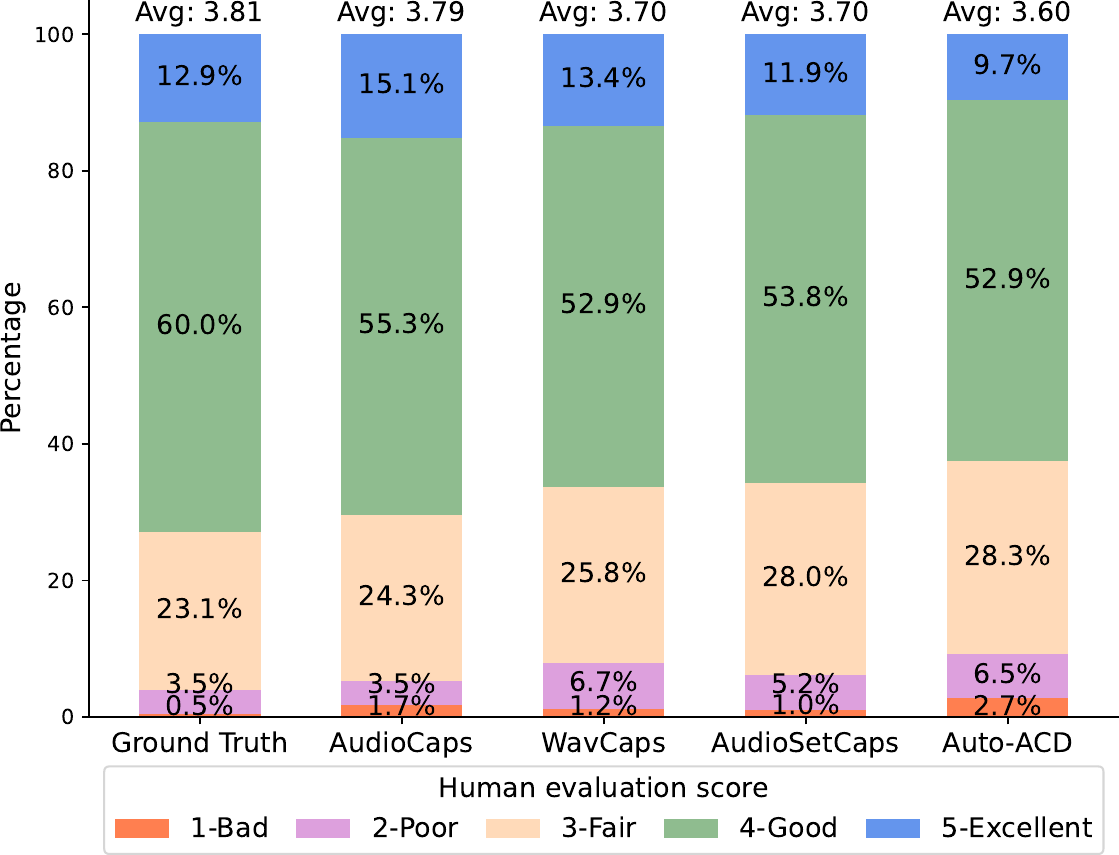}
    \caption{Mean opinion score distribution for ground truth labels and captions of each dataset. Each score ranges from $1$ (Bad) to $5$ (Excellent), with the average score shown above each bar. The evaluation is conducted on $79$ overlapping audio samples with at least $5$ evaluators per sample.}
    \vspace{-5pt}
    \label{fig:MOS_per}
\end{figure}

To further assess the quality of AudioSetCaps, we perform subjective evaluation\footnote{Our subjective test with Amazon Mechanical Turk has received a favourable opinion from the ethics review after completing the University of Surrey Self-Assessment Governance and Ethics form under Application No. 1046015-1045997-121266365.} to evaluate captions corresponding to $79$ overlapping audio samples across AudioCaps \cite{kim2019audiocaps}, WavCaps~\cite{mei2024wavcaps}, AudioSetCaps, and Auto-ACD \cite{sun2024auto}, as well as the ground truth labels in AudioSet~\cite{gemmeke2017audio}.
The question we ask for the rater is \textit{Please listen to the provided audio samples and rate the quality of the text annotation based on its accuracy, completeness, and presence of false information.}
Each caption or label is scored by at least five evaluators on how well the text annotation reflects the audio content, ranging from $1$ to $5$, where $1$ represents ``Bad'', and $5$ represents ``Excellent''. 

The mean opinion score (MOS) and distributions for each dataset and the ground truth labels are presented in Fig.~\ref{fig:MOS_per}.
AudioSetCaps achieves a MOS of $3.70$, which closely approaches the quality of both the ground truth labels and human-labelled captions from AudioCaps. 
Ground truth achieves the highest MOS of $3.81$, with $60.0$\% rated as ``Good'' and $12.9$\% as ``Excellent''. 
AudioCaps achieves a MOS of $3.79$, showing the highest percentage of ``Excellent'' ratings at $15.1$\%. 
WavCaps and AudioSetCaps both have a MOS of $3.70$. 
Their score distributions are similar, where AudioSetCaps shows a slightly higher percentage of ``Fair" ratings at $25.8$\% and a lower percentage of ``Poor" ratings at $5.2$\%. 
Auto-ACD has a MOS of $3.60$ with the lowest percentage of ``Excellent'' ratings but the highest percentage of ``Fair'' ratings. 
The automatically generated datasets, especially WavCaps and AudioSetCaps, perform comparably to human-annotated datasets.

\section{Experiments}
\label{sec:exps}

To evaluate the proposed AudioSetCaps dataset, we conducted experiments on several audio-language tasks, including AAC, ATR, and zero-shot audio classification. 
Through the experiments, we aim to validate both the quality of AudioSetCaps and the effectiveness of the dataset for downstream tasks. 
Moreover, we provide a comprehensive analysis of the tasks including the training data scaling study, ablation study on the proposed pipeline, and a detailed case study of captions in different datasets. 

\subsection{Automated Audio Captioning}
Automated audio captioning~\cite{drossos2017automated, mei2022automated, xu2023beyond} is a cross-modal task that bridges audio understanding and natural language generation. An AAC system primarily employs an encoder-decoder neural architecture, where the encoder processes and extracts acoustic features from audio signals while the decoder generates natural language descriptions. 
The AAC model can be trained by a cross-entropy loss:
\begin{equation}
\mathcal{L}_{AAC}=-\frac{1}{N}\sum_{n=1}^{N}\log p({c}_{n}| {c}_{1:n-1},x)
\end{equation}
where $x$ is an input audio clip, ${c}_{n}$ is the $n$-th ground truth token in a sentence with a length of $N$.

\subsubsection{Model and Experimental Settings}
The AAC model used in our experiments follows the architecture used in WavCaps~\cite{mei2024wavcaps}, utilizing pre-trained HTSAT~\cite{chen2022hts} as the audio encoder and BART~\cite{lewis-etal-2020-bart} as the text decoder. 
For training, we pre-train the model on either AudioSetCaps alone or in combination with Clotho and AudioCaps training sets using a learning rate of $5\times 10^{-5}$ and a batch size of $24$ for $15$ epochs. 
The model is then fine-tuned on AudioCaps for $20$ epochs with a learning rate of $5\times 10^{-6}$.
Note that the validation and test sets of AudioCaps are excluded during pre-training and fine-tuning. 
The performance is evaluated using the conventional AAC metrics including BLEU$_{n}$~\cite{papineni2002bleu}, ROUGE$_{l}$~\cite{lin2004rouge}, METEOR~\cite{banerjee2005meteor}, CIDEr~\cite{vedantam2015cider}, SPICE~\cite{anderson2016spice} and SPIDEr~\cite{liu2017improved}.

\begin{table*}[t]
\centering
\caption{The performance of state-of-the-art methods for automated audio captioning on AudioCaps test set.}
\begin{tabular}{cccccccccc}
\toprule
\textbf{Method}                        & \textbf{Training Set}                  & \textbf{Model}                                 & \textbf{BLEU$_{1}$}         & \textbf{BLEU$_{4}$}         & \textbf{ROUGE$_{l}$}        & \textbf{METEOR}        & \textbf{CIDEr}         & \textbf{SPICE}         & \textbf{SPIDEr}        \\ \midrule
Qwen-Audio~\cite{chu2023qwen}                   & N/A                           & Wisper+Qwen 7B                        & -             & -             & -          & -             & -             & $14.7$             & -             \\
Pengi~\cite{deshmukh2023pengi}                         & AC+CL              & HTSAT+CLIP+GPT-2                   & $69.1$          & $25.3$          & $48.2$          & $23.2$          & $75.2$          & $18.2$          & $46.7$          \\
AL-MixGen~\cite{liu2017improved}                     & AC                     & ACT             & $70.0$          & $28.9$          & $50.2$          & $24.2$          & $76.9$          & $18.1$          & $47.5$          \\
\multirow{2}{*}{EnCLAP~\cite{kim2024enclap}}       & AC                     & EnCLAP-base                           & -             & -             & -             & $24.7$          & $78.0$          & $18.6$          & $48.3$          \\
                              & AC                     & EnCLAP-large                          & -             & -             & -             & $25.5$          & $80.3$          & $18.8$          & $49.5$          \\
\multirow{2}{*}{CoNeTTE~\cite{labb2024conette}}      & AudioSet               & \multirow{2}{*}{CNeXt-Trans} & -             & -             & -             & -             & -             & -             & $49.5$          \\
                              & AudioSet+AC            &                                       & -             & -             & -             & -             & -             & -             & $46.6$          \\
AutoCap~\cite{haji2024taming}  & \multirow{2}{*}{AC+CL+WavCaps} & CLAP+HTSAT+  & \multirow{2}{*}{$72.3$}   & \multirow{2}{*}{$29.7$}  & \multirow{2}{*}{$51.8$}  & \multirow{2}{*}{$25.3$}  & \multirow{2}{*}{$83.2$} & \multirow{2}{*}{$18.2$} & \multirow{2}{*}{$50.7$}  \\
                            &                          & Q-Former+BART                   &          &           &           &           &           &           &           \\
\multirow{2}{*}{WavCaps~\cite{mei2024wavcaps}}      & AC+CL              & \multirow{2}{*}{HTSAT+BART}           & $67.5$          & $27.2$          & $48.3$          & $23.7$          & $71.1$          & $17.7$          & $44.4$          \\
                              & WavCaps+AC+CL      &                                       & $70.7$          & $28.3$          & $50.7$          & $25.0$          & $78.7$          & $18.2$          & $48.5$          \\ \midrule
\multirow{2}{*}{AudioSetCaps} & AudioSetCaps+AC+CL & \multirow{2}{*}{HTSAT+BART}           & $71.4$          & $30.3$          & $52.2$          & $25.7$          & $\mathbf{84.8}$ & $18.4$          & $\mathbf{51.6}$ \\ 
                              & AudioSetCaps                  &                                       & $\mathbf{73.9}$ & $\mathbf{30.8}$ & $\mathbf{52.7}$ & $\mathbf{26.2}$ & $83.9$          & $\mathbf{18.6}$ & $51.3$          \\ \bottomrule
\end{tabular}
\label{tab:aac_results}
\begin{tablenotes}
\item ``AC'' and ``CL'' refer to AudioCaps and Clotho datasets. ``ACT'' refers to the audio captioning transformer model.
\end{tablenotes}
\end{table*}

\subsubsection{Results}
Table~\ref{tab:aac_results} presents the AAC results on the AudioCaps test set. 
The experimental results demonstrate the promising performance of the HTSAT and BART model trained on AudioSetCaps for AAC. 
The AAC model trained on the combination of AudioSetCaps, AudioCaps, and Clotho not only surpasses the performance of Qwen-Audio by a large margin but also outperforms WavCaps and many AAC methods such as Pengi~\cite{deshmukh2023pengi}, AL-MixGen~\cite{liu2017improved}, EnCLAP~\cite{kim2024enclap}, and CoNeTTE~\cite{labb2024conette}. 
Notably, the model trained solely on AudioSetCaps further improves the AAC performance on several metrics, outperforming the state-of-the-art AutoCap~\cite{haji2024taming} across all metrics.

The comparison between models trained solely on AudioSetCaps and with combined datasets reveals an interesting pattern in performance metrics. 
While the model trained on combined datasets shows slightly higher performance in CIDEr and SPIDEr scores, the model trained exclusively on AudioSetCaps achieves better results in BLEU, ROUGE$_{l}$, METEOR, and SPICE metrics. 
These results suggest that AudioSetCaps contains abundant audio characteristics for caption generation, and the incorporation of additional datasets does not lead to consistent improvements across all metrics.

\subsection{Audio-Text Retrieval}

\begin{table*}[t]
\centering
\caption{Performance of state-of-the-art methods for audio-text retrieval on the AudioCaps test set. Note that the Multilingual method employs a multilingual language-enhanced approach, and its performance is provided here for reference only, not for direct comparison.}
\begin{tabular}{ccccccccc}
\toprule
\multirow{2}{*}{\textbf{Method}}       & \multirow{2}{*}{\textbf{Training Set}}       & \multirow{2}{*}{\textbf{Model}} & \multicolumn{3}{c}{\textbf{Text-to-Audio}}                       & \multicolumn{3}{c}{\textbf{Audio-to-Text}}                       \\ \cmidrule(lr){4-6} \cmidrule(lr){7-9}
                              &                                     &                        & \textbf{R@1}            & \textbf{R@5}            & \textbf{R@10}           & \textbf{R@1}            & \textbf{R@5}            & \textbf{R@10}           \\ \midrule
{\color{gray}Multilingual~\cite{yan2024bridging}}           & {\color{gray}WavCaps+AC+CL}         & {\color{gray}CED-BASE+SONAR (LE)}     & {\color{gray}$45.9$}          & {\color{gray}$\mathbf{81.3}$} & {\color{gray}$\mathbf{90.2}$} & {\color{gray}$\mathbf{60.7}$} & {\color{gray}$\mathbf{86.9}$} & {\color{gray}$\mathbf{94.8}$} \\ \midrule
LAION CLAP~\cite{wu2023large}                        & LA+AudioSet+AC+CL                   & HTSAT+RoBERTa          & $36.1$          & $71.8$          & $83.9$          & $46.8$          & $82.9$          & $90.7$          \\
\multirow{2}{*}{WavCaps~\cite{mei2024wavcaps}}      & AC+CL                               & HTSAT+BERT             & $39.2$          & $74.9$          & $86.5$          & $49.5$          & $81.9$          & $91.5$          \\
                              & WavCaps+AC+CL                       & HTSAT+BERT             & $42.2$          & $76.5$          & $87.1$          & $54.6$          & $85.2$          & $92.4$          \\
Auto-ACD~\cite{sun2024auto}                      & Auto-ACD+AC+CL                      & HTSAT+RoBERTa (TM)      & $42.7$          & -             & $88.5$          & $56.3$          & -             & $\mathbf{93.9}$ \\
\multirow{2}{*}{Sound-VECaps~\cite{yuan2024improving}} & Sound-VECapsF                       & HTSAT+RoBERTa          & $39.2$          & $74.1$          & $85.0$          & $54.0$          & $85.5$          & $93.2$          \\
                              & Sound-VECapsA                       & HTSAT+RoBERTa          & $41.2$          & $74.5$          & $85.3$          & $53.3$          & $83.2$          & $93.0$          \\ \midrule
\multirow{4}{*}{AudioSetCaps} & \multirow{2}{*}{AudioSetCaps+AC+CL} & HTSAT+BERT             & $43.4$          & $78.4$          & $88.2$          & $57.3$          & $84.2$          & $93.2$          \\
                              &                                     & HTSAT+RoBERTa (TM)      & $44.8$          & $79.2$          & $88.8$          & $59.6$          & $85.1$          & $92.2$          \\
                              & \multirow{2}{*}{AudioSetCaps}       & HTSAT+RoBERTa          & $44.4$          & $79.3$          & $89.6$          & $58.6$          & $\mathbf{86.2}$ & $\mathbf{93.9}$ \\
                              &                                     & HTSAT+RoBERTa (TM)      & $\mathbf{46.3}$ & $\mathbf{79.7}$ & $\mathbf{90.0}$ & $\mathbf{59.7}$ & $86.0$          & $93.2$          \\ \bottomrule
\end{tabular}
\label{tab:atr_results}
\begin{tablenotes}
\item ``LA'', ``AC'' and ``CL'' refer to LAION-Audio-630K, AudioCaps and Clotho datasets. ``TM'' denotes the text masking method during training. ``R@$k$" denotes recall at rank $k$. 
\end{tablenotes}
\end{table*}

Audio-text retrieval ~\cite{Xie2022Language, mei22onmetric, lou2022audio} is a cross-modal task that aims to address the bidirectional retrieval between acoustic signals and natural language descriptions. 
For example, the ATR system can retrieve the most relevant text for an audio query or the best-matched audio for a text query from a candidate pool.
Recent ATR systems typically adopt a contrastive learning framework, where the model learns to maximize the similarity between paired audio-text embeddings while minimizing the unpaired ones. 
The audio infoNCE loss~\cite{he2020momentum}, $\mathcal{L}_{a}$, can be defined as the average of a normalized function measuring the similarity of different texts to the same audio query. 
Similarly, the loss for the text modality, $\mathcal{L}_{t}$, measures the similarity of different audios to the same text query. 
For a batch with $B$ audio-text pairs, we have:
\begin{equation}
\label{eq:1}
\mathcal{L}_{i}^{a}=-\log \frac{\exp \left(e_{i}^{a} \cdot e_{i}^{t} / \tau\right)}{\sum_{j=1}^{B} \exp \left(e_{i}^{a} \cdot e_{j}^{t} / \tau\right)}
\end{equation}
\begin{equation}
\label{eq:2}
\mathcal{L}_{i}^{t}=-\log \frac{\exp \left(e_{i}^{t} \cdot e_{i}^{a} / \tau\right)}{\sum_{j=1}^{B} \exp \left(e_{i}^{t} \cdot e_{j}^{a} / \tau\right)}
\end{equation}
where $\tau$ is a temperature parameter that controls the range of logits, $e_{i}^{a}$ and $e_{i}^{t}$ are the embedding vectors of the $i$-th audio sample $x_{i}$ and text caption $c_{i}$ in a training batch, respectively. 
The loss $\mathcal{L}_{ATR}$ for the audio-text pairs in one batch can be defined as a symmetrical objective:
\begin{equation}
\label{eq:3}
\mathcal{L}_{ATR}=\frac{1}{2B} \sum_{i=1}^{B}(\mathcal{L}_{i}^{a} + \mathcal{L}_{i}^{t})
\end{equation}

\subsubsection{Model and Experimental Settings}
We implement the ATR model following WavCaps~\cite{mei2024wavcaps}, which consists of three components: a pre-trained HTSAT~\cite{chen2022hts} for audio encoding, pre-trained BERT-base networks~\cite{devlin2019bert, liu2019roberta} for text encoding, and a 2-layer multi-layer perceptron with ReLU activation for feature alignment. 
The model is first pre-trained either on AudioSetCaps alone or on a combined dataset of AudioSetCaps, AudioCaps, and Clotho for 40 epochs with a batch size of 196 and a learning rate of $5\times 10^{-5}$. 
For supervised fine-tuning, the model is further trained on AudioCaps for 20 epochs with a learning rate of $1\times 10^{-5}$. 
We evaluate the model performance using recall at rank $k$ (R@$k$), which measures the percentage of queries that retrieve the correct item within the top $k$ results.

\subsubsection{Main Results}
Table~\ref{tab:atr_results} presents the ATR results on the AudioCaps test dataset. 
The experimental results for the ATR model trained on AudioSetCaps demonstrate state-of-the-art performance across the metrics.
Compared with LAION CLAP~\cite{wu2023large}, the AudioSetCaps-based models show superior results.
Our primary comparison focuses on other LLM-based automated audio caption generation pipelines, such as WavCaps, Auto-ACD, and Sound-VECaps. 
We first train an ATR model using HTSAT and BERT on the combination of AudioSetCaps, AudioCaps, and Clotho. 
This model achieves an R@1 score of $43.4$ for text-to-audio retrieval and an R@1 score of $57.3$ for audio-to-text retrieval, surpassing the model trained on the combination of WavCaps, AudioCaps, and Clotho. 
When we employ HTSAT with RoBERTa ~\cite{liu2019roberta} and the text masking method, the ATR model yields R@1 scores of $44.8$ and $59.6$ for text-to-audio and audio-to-text retrieval, respectively, outperforming the same ATR model used in Auto-ACD~\cite{sun2024auto}. 
The HTSAT with RoBERTa model trained solely on AudioSetCaps attains R@1 scores of $44.4$ and $58.6$ for text-to-audio and audio-to-text retrieval, surpassing the Sound-VECaps approach. 
Furthermore, the HTSAT with RoBERTa model achieves improved performance when applying text masking, reaching R@1 scores of $46.3$ and $59.7$ for text-to-audio and audio-to-text retrieval, respectively. 
These results approach the performance of the state-of-the-art multilingual method~\cite{yan2024bridging} which uses a multilingual language-enhanced ATR framework with CED-BASE \cite{dinkel2024ced} and SONAR models~\cite{duquenne2023sonar}.

\subsubsection{Ablation Study for ATR Models}

\begin{table}[t]
\centering
\caption{Ablation study of different pre-training datasets, text masking ratios, and text encoders for audio-text retrieval models on the AudioCaps test set.}
\resizebox{\columnwidth}{!}{%
\begin{tabular}{cccccc}
\toprule
\multirow{2}{*}{\textbf{Model}} & \multicolumn{2}{c}{\textbf{Training Set}} & \textbf{Mask} & \textbf{T2A} & \textbf{A2T} \\
\cmidrule(lr){2-3} \cmidrule(lr){5-6}
& \textbf{ASC} & \textbf{ASC+AC+CL} & (\%) & \textbf{R@1} & \textbf{R@1} \\
\cmidrule(lr){1-6}
\multirow{2}{*}{HTSAT+RoBERTa-PT} & $\times$ & $\checkmark$ & - & $40.3$ & $54.4$ \\
       & $\checkmark$ & $\times$ & - & $30.8$ & $41.4$ \\
\midrule
\multirow{2}{*}{HTSAT+RoBERTa-FT} & $\times$ & $\checkmark$ & - & $42.9$ & $56.5$ \\ 
& $\checkmark$ & $\times$ & - & $44.4$ & $58.6$ \\  \cmidrule(lr){1-6}
\multirow{2}{*}{HTSAT+RoBERTa-FT} & $\checkmark$ & $\times$ & $25$ & $\mathbf{46.3}$ & $\mathbf{59.7}$ \\ 
& $\checkmark$ & $\times$ & $50$ & $43.4$ & $56.4$ \\ 
\cmidrule(lr){1-6}
HTSAT+BERT-FT & $\checkmark$ & $\times$ & - & $44.8$ & $55.7$ \\
\bottomrule
\end{tabular}
}
\label{tab:atr_ablation}
\begin{tablenotes}
\item ``ASC'' refers to AudioSetCaps, ``AC'' refers to AudioCaps, ``CL'' refers to Clotho, ``T2A'' refers text-to-audio retrieval, ``A2T'' refers audio-to-text retrieval, and ``R@1'' refers to recall at rank 1.
\end{tablenotes}
\end{table}

We further conduct ablation studies to analyze the impact of using different pre-training datasets, text masking ratios, and text encoders for training ATR models. 
Experimental results on AudioCaps test set are shown in Table~\ref{tab:atr_ablation}.

\textit{Pre-training Dataset}: We study the impact of different pre-training datasets on the ATR performance using the ``HTSAT+RoBERTa'' model. 
Initially, the model is pre-trained using two configurations: solely AudioSetCaps, and a combination of AudioSetCaps, AudioCaps, and Clotho. 
The results show that pre-training on the combined dataset yields better initial performance. 
However, after fine-tuning on AudioCaps, the model pre-trained solely on AudioSetCaps achieves better performance in text-to-audio and audio-to-text retrieval tasks. 
This suggests that while the combined dataset can provide a better starting point, the pre-trained model solely AudioSetCaps can bridge the performance gap after the fine-tuning process on a targeted dataset like AudioCaps.
Similar results can be observed from the AAC experiments that AudioSetCaps is a good audio-language learning dataset with diverse and extensive audio content.

\textit{Text Masking Ratio}: We investigate the effect of different text masking ratios on the ``HTSAT+RoBERTa" model pre-trained on AudioSetCaps. 
Three masking configurations are examined, i.e., no masking, $25$\% masking, and 50\% masking. 
The experimental results show that $25$\% text masking effectively improves the ATR performance, with R@1 scores of $46.3$\% for text-to-audio retrieval and $59.7$\% for audio-to-text retrieval. 
Similarly, the ATR model in Auto-ACD also adopts $25$\% text masking \cite{sun2024auto}. 
Excessive masking, such as 50\%, degrades the performance, suggesting the importance of maintaining a balance between introducing robustness and preserving semantic information during training.

\textit{Text Encoder}: We study the impact of using different text encoders in ATR models. 
While the BERT model achieves a slightly higher R@1 score of $44.8$\% for text-to-audio retrieval compared to $44.4$\% of RoBERTa, RoBERTa demonstrates much better performance for audio-to-text retrieval. 
The results indicate that RoBERTa serves as a more effective text encoder for the ATR task.

\begin{table*}[t!]
\centering
\caption{Performance comparison of zero-shot audio classification for environmental sound, speech, and music tasks.}
\begin{tabular}{cccccccc}
\toprule
\multirow{3}{*}{\textbf{Method}} & \multicolumn{2}{c}{\textbf{Environmental Sound}}                                 & \multicolumn{3}{c}{\textbf{Speech}}                           & \multicolumn{2}{c}{\textbf{Music}}        \\ \cmidrule(lr){2-3} \cmidrule(lr){4-6} \cmidrule(lr){7-8}
                        & \multicolumn{2}{c}{Sound Classification} & Spoken Language & \multicolumn{2}{c}{Speech Emotion} & Music Genre  & Musical Instrument \\ \cmidrule(lr){2-2} \cmidrule(lr){3-3} \cmidrule(lr){4-4} \cmidrule(lr){5-6} \cmidrule(lr){7-7} \cmidrule(lr){8-8}
                        & UrbanSound8K                  & ESC-50                        & CommonLanguage  & CREMA-D          & RAVDESS         & GTZAN-Genre   & OpenMIC-2018     \\ \midrule
LAION CLAP~\cite{wu2023large}              & $77.0$                            & $91.0$                      & $15.5$            & $18.6$             & $11.5$            & $43.0$            & $54.0$             \\
WavCaps~\cite{mei2024wavcaps} & $\mathbf{80.6}$                 & $\mathbf{94.8}$           & $11.8$            & $20.1$             & $14.5$            & $44.3$          & $46.2$            \\
Auto-ACD~\cite{sun2024auto}                & $76.2$                          & $86.1$                    & $13.9$            & $20.2$             & $10.8$            & $45.6$          & $54.4$               \\ \midrule
Ours            & $76.6$                          & $88.0$                      & $\mathbf{32.6}$   & $\mathbf{28.5}$    & $\mathbf{25.7}$   & $\mathbf{70.5}$ & $\mathbf{59.7}$    \\ \bottomrule
\end{tabular}
\label{tab:zsc_results}
\end{table*}

\subsection{Zero-Shot Audio Classification}

Zero-shot audio classification~\cite{xie2019zero,xie2021zero} is a task that can recognize unseen sound categories without further training. 
Leveraging the capabilities of ATR models, zero-shot audio classification generalizes to unseen classes by treating class labels or descriptions as text queries. 
By calculating the semantic similarity between these queries and audio embeddings within a shared embedding space, the model predicts the class with the highest similarity score as the category.

We employ the ``HTSAT+RoBERTa" ATR model pre-trained on the combination of AudioSetCaps, AudioCaps, and Clotho datasets with $25$\% text masking, to study the zero-shot performance on audio classification tasks. 
We extend the labels to sentences and use them as the input text of the ATR model following the input text in WavCaps~\cite{mei2024wavcaps}.
This model is evaluated on seven diverse audio classification tasks across the following three categories.

\noindent
\textbf{Environmental Sound}: UrbanSound8K~\cite{salamon2014dataset} ($8,732$ samples, $10$ classes) and ESC-50~\cite{piczak2015esc} ($2,000$ samples, 50 classes).

\noindent
\textbf{Speech}: CommonLanguage (a subset of CommonVoice~\cite{ardila2020common}) for language recognition ($978$ samples, $6$ languages), CREMA-D~\cite{cao2014crema} ($7,442$ samples, $6$ emotions) and RAVDESS~\cite{livingstone2018ryerson} ($2,452$ samples, $8$ emotions) for emotion recognition.

\noindent
\textbf{Music}: GTZAN-Genre~\cite{tzanetakis_essl_cook_2001} for genre classification ($1,000$ samples, $10$ genres) and OpenMIC-2018~\cite{Humphrey2018OpenMIC2018AO} for instrument recognition ($13,847$ samples, $20$ instruments).

Results of zero-shot audio classification on environmental sound, speech, and music datasets are shown in Table \ref{tab:zsc_results}.
The ATR model trained on AudioSetCaps demonstrates strong performance across various audio classification tasks, especially on speech and music tasks. 
In environmental sound classification, the model achieved competitive results with $76.6$\% accuracy on UrbanSound8K and $88.0$\% on ESC-50, closely approaching the performance of WavCaps trained with diverse audio sources. 
For language identification using CommonLanguage, it significantly outperformed other approaches with an accuracy of $32.6$\%. 
In speech emotion recognition, the model achieved better accuracies of $28.5$\% on CREMA-D and $25.7$\% on RAVDESS. 
For music genre classification using GTZAN-Genre, it achieves an accuracy of $70.5$\%, substantially outperforming other methods. 
Similarly, in musical instrument recognition on OpenMIC-2018, the model attains the highest accuracy of $59.7$\%.
These results demonstrate the effectiveness of incorporating diverse fine-grained audio content into the caption, potentially improving the generalized audio understanding for audio-language models.


\begin{figure}[t]
    \centering
    \includegraphics[width=1\linewidth]{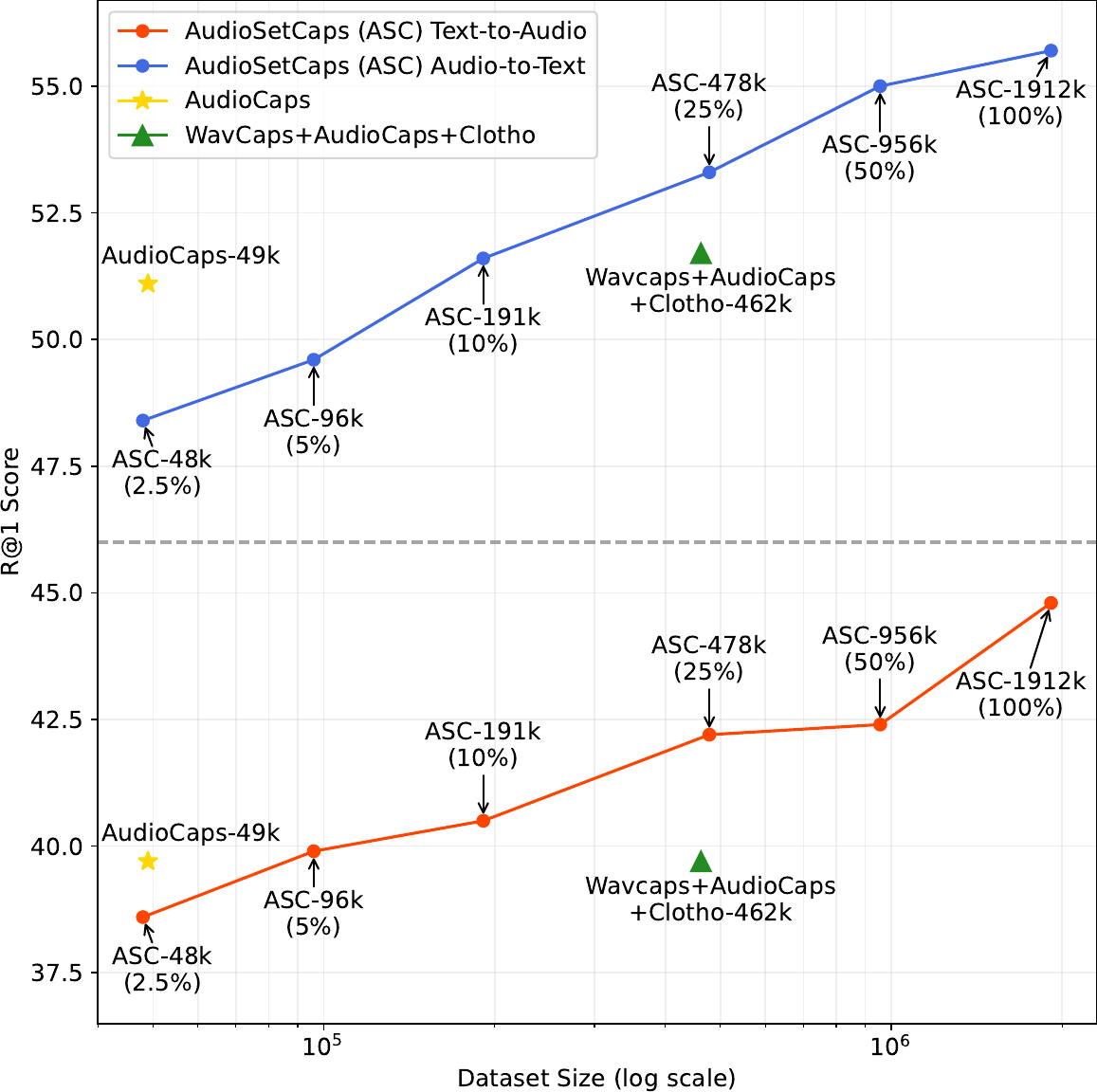}
    \caption{Data scaling performance of AudioSetCaps on audio-text retrieval.}
    \label{fig:ATR_scaling}
\end{figure}

\subsection{Training Data Scaling Study}

To study the impact of using different data volumes of AudioSetCaps for training, we conducted experiments using the HTSAT and BERT models for ATR tasks. 
We created six pre-defined ASC subsets: $100$\% ($1912$K), $50$\% ($956$K), $25$\% ($478$K), $10$\% ($191$K), $5$\% ($96$K), and $2.5$\% ($48$K), where the smaller subsets are fully included in the larger ones. 
By maintaining consistency across subsets, we can directly observe how model performance scales with increasing data volume. 
This setup also enables us to identify potential inflection points in the learning curve, where additional data may yield improved performance, providing insights into the data efficiency for audio-language learning. For fair comparisons, the subset of $478$K ($25$\%) is comparable in size to the combination of WavCaps, AudioCaps, and Clotho ($462$K), and the subset of $48$K ($2.5$\%) is similar to AudioCaps ($49$K).

For the audio-to-text results, AudioSetCaps demonstrates consistent performance improvements as the training data scales up. 
Notably, with just $10$\% of its data ($191$K samples), AudioSetCaps already matches the performance of AudioCaps and ``WavCaps+AC+Clotho''. 
For the text-to-audio results, we observe a similar trend to the audio-to-text results of AudioSetCaps. 
The model trained on $25$\% of AudioSetCaps ($478$K samples) surpasses both AudioCaps and the combination of WavCaps, AudioCaps, and Clotho. 
These results suggest that AudioSetCaps provides enriched and accurate audio content for training ATR models, demonstrating effectiveness in further promoting the audio-language learning tasks.

\subsection{Ablation Study on Pipeline}

The ablation study of the AudioSetCaps pipeline aims to evaluate the effectiveness of key components in our automated caption generation process. 
We investigate three critical aspects of our pipeline: the inclusion of fine-grained speech and music content, prompt chaining for Qwen-Audio, and the caption refinement step. 
We use the ``HTSAT+BERT'' in the experiments as the baseline ATR model while using different training sets. 
We train the ATR models using the AudioCaps captions and those with the same audio ID in AudioCaps but generated by the AudioSetCaps pipeline.
We alternately remove each component to assess its impact on the performance. Experimental results are shown in Table \ref{tab:pipeline_ablation}.

\begin{table}[t]
\centering
\caption{Ablation study of components in the proposed audio caption generation pipeline for audio-text retrieval tasks on the AudioCaps test set.}
\begin{tabular}{lcc}
\toprule
\multirow{2}{*}{\textbf{Settings}}                   & \textbf{T2A} & \textbf{A2T} \\ \cmidrule(lr){2-3}
                             & \textbf{R@1}     & \textbf{R@1}    \\  \midrule
AudioCaps                & $39.7$             & $51.1$             \\                          
AudioCaps (AudioSetCaps pipeline)            & $39.7$             & $50.9$             \\ \midrule
\quad w/o Fine-grained speech and music content          & $38.4$             & $48.3$             \\
\quad w/o Prompt chaining for Qwen-Audio             & $39.6$             & $49.2$             \\
\quad w/o Caption refinement              & $38.7$             & $48.9$                  \\ \bottomrule
\end{tabular}
\begin{tablenotes}
\item ``T2A'' refers text-to-audio retrieval, ``A2T'' refers audio-to-text retrieval, and ``R@1'' refers to recall at rank 1.
\end{tablenotes}
\label{tab:pipeline_ablation}
\end{table}

The use of AudioSetCaps pipeline-generated captions as the training dataset achieves results comparable to those of the AudioCaps dataset. 
We then remove the fine-grained speech and music content within the pipeline, which leads to the most significant drop in performance. 
Next, we replace the prompt chaining by merging all the prompts for Qwen-Audio, resulting in decreased scores of 39.6\% and 49.2\% on text-to-audio and audio-to-text retrieval, respectively. 
Finally, we skip the caption refinement step during the caption generation, which also indicates a decrease in the ATR performance. 
These findings suggest that each component of our automated pipeline is important, and the proposed pipeline can generate captions of quality approaching human annotations.

\subsection{Extended Audio-Language Datasets for Future Work}

\begin{table}[t]
\centering
\caption{Statistics for large-scale audio-language datasets generated using the proposed pipeline. QA denotes question-answer.}
\begin{tabular}{lrr|r}
\toprule
\textbf{Dataset} & \textbf{Caption data} &\textbf{QA data} & \textbf{Total} \\
\midrule
\textbf{AudioSetCaps} & $1,910,920$ & $5,736,072$ & $7,646,992$ \\
\textbf{YouTube-8M} & $4,023,990$ & $12,086,037$ & $16,110,027$ \\
\textbf{VGGSound} & $182,189$ & $592,680$ & $774,869$ \\
\midrule
\textbf{Total} & $6,117,099$ & $18,414,789$ & $\mathbf{24,531,888}$ \\
\bottomrule
\end{tabular}
\label{tab:extended_dataset_quantity}
\end{table}

As demonstrated by the result shown in Table~\ref{tab:aac_results}, Table~\ref{tab:atr_results}, Table~\ref{tab:zsc_results}, and Fig.~\ref{fig:ATR_scaling}, AudioSetCaps significantly benefit the audio-language learning tasks and shows a strong correlation between performance gain and dataset scaling. This indicates our proposed synthetic audio caption generation pipeline works well for building large-scale audio-language datasets.
To explore the boundary of data scaling and facilitate future research, We extend the proposed pipeline to generate audio-caption data for YouTube-8M and VGGSound datasets, 
creating over $6$~M audio captions and $18$~M audio QA pairs across three datasets. 
The statistics of our collected audio caption and QA data are summarized in Table \ref{tab:extended_dataset_quantity}.
We also make this extended audio-language dataset available to the public and will treat the experiment on this extended dataset as our future work.

\section{Conclusion}
\label{sec:Conclusion}
This paper has proposed an automated pipeline that leverages large audio and language models to generate enriched audio-language data. 
Our pipeline comprises three key stages: LALMs-driven audio content extraction, LLMs-assisted caption generation, and CLAP model-based refinement. 
Using this pipeline, we created AudioSetCaps, containing $1.9$M audio-caption pairs with fine-grained speech and music contents.
Extensive experiments demonstrated the effectiveness of our approach. 
Models trained on AudioSetCaps achieved state-of-the-art performance on AAC, ATR, and zero-shot audio classification tasks. 
Our ablation studies verified the importance of each pipeline component, while the data scaling analysis showed that even a subset of AudioSetCaps can provide competitive performance compared to existing datasets.



\bibliography{refers}
\bibliographystyle{IEEEtran}

\end{document}